\documentclass[11pt,a4paper]{article}

\usepackage[margin=2.5cm]{geometry}

\usepackage{amsmath,amssymb}
\usepackage{graphicx}
\usepackage{booktabs}
\usepackage{longtable}
\usepackage{array}
\usepackage{multirow}
\usepackage{xcolor}
\usepackage{hyperref}
\usepackage{xurl}
\usepackage{listings}
\usepackage{microtype}
\usepackage[htt]{hyphenat}

\usepackage{iftex}
\ifPDFTeX
  \usepackage[T1]{fontenc}
  \usepackage[utf8]{inputenc}
  \usepackage{textcomp} 
\else 
  \usepackage{unicode-math} 
  \defaultfontfeatures{Scale=MatchLowercase}
  \defaultfontfeatures[\rmfamily]{Ligatures=TeX,Scale=1}
\fi
\usepackage{lmodern}
\ifPDFTeX\else
\fi
\IfFileExists{upquote.sty}{\usepackage{upquote}}{}
\IfFileExists{microtype.sty}{
  \usepackage[]{microtype}
  \UseMicrotypeSet[protrusion]{basicmath} 
}{}
\makeatletter
\@ifundefined{KOMAClassName}{
  \IfFileExists{parskip.sty}{%
    \usepackage{parskip}
  }{
    \setlength{\parindent}{0pt}
    \setlength{\parskip}{6pt plus 2pt minus 1pt}}
}{
  \KOMAoptions{parskip=half}}
\makeatother
\usepackage{color}
\usepackage{fancyvrb}

\DefineVerbatimEnvironment{Highlighting}{Verbatim}{commandchars=\\\{\}}
\newenvironment{Shaded}{}{}

\newcommand{\AttributeTok}[1]{\textcolor[rgb]{0.49,0.56,0.16}{#1}}

\newcommand{\DecValTok}[1]{\textcolor[rgb]{0.25,0.63,0.44}{#1}}

\newcommand{\FunctionTok}[1]{\textcolor[rgb]{0.02,0.16,0.49}{#1}}

\newcommand{\KeywordTok}[1]{\textcolor[rgb]{0.00,0.44,0.13}{\textbf{#1}}}
\newcommand{\NormalTok}[1]{#1}

\newcommand{\StringTok}[1]{\textcolor[rgb]{0.25,0.44,0.63}{#1}}
\newcommand{\VariableTok}[1]{\textcolor[rgb]{0.10,0.09,0.49}{#1}}

\usepackage{longtable,booktabs,array}
\usepackage{calc} 
\usepackage{etoolbox}
\makeatletter
\patchcmd\longtable{\par}{\if@noskipsec\mbox{}\fi\par}{}{}
\makeatother
\IfFileExists{footnotehyper.sty}{\usepackage{footnotehyper}}{\usepackage{footnote}}
\makesavenoteenv{longtable}
\NewDocumentCommand\citeproctext{}{}

\makeatletter
 \let\@cite@ofmt\@firstofone
 \def\@biblabel#1{}
 \def\@cite#1#2{{#1\if@tempswa , #2\fi}}
\makeatother
\newlength{\cslhangindent}
\newlength{\csllabelwidth}
\newenvironment{CSLReferences}[2] 
 {\begin{list}{}{%
  \setlength{\itemindent}{0pt}
  \setlength{\leftmargin}{0pt}
  \setlength{\parsep}{0pt}
  \ifodd #1
   \setlength{\leftmargin}{\cslhangindent}
   \setlength{\itemindent}{-1\cslhangindent}
  \fi
  \setlength{\itemsep}{#2\baselineskip}}}
 {\end{list}}
\usepackage{calc}

\providecommand{\tightlist}{%
  \setlength{\itemsep}{0pt}\setlength{\parskip}{0pt}}
\makeatletter
\@ifpackageloaded{subcaption}{}{\usepackage{subcaption}}
\@ifpackageloaded{caption}{}{\usepackage{caption}}
\AtBeginDocument{%

}
\AtBeginDocument{%

}
\newcounter{pandoccrossref@subfigures@footnote@counter}
{\end{figure}%
\addtocounter{footnote}{-\value{pandoccrossref@subfigures@footnote@counter}}
\@for\f:=\global@pandoccrossref@subfigures@footnotes\do{\stepcounter{footnote}\footnotetext{\f}}%
\gdef\global@pandoccrossref@subfigures@footnotes{}}
\@ifpackageloaded{float}{}{\usepackage{float}}
\floatstyle{ruled}
\@ifundefined{c@chapter}{\newfloat{codelisting}{h}{lop}}{\newfloat{codelisting}{h}{lop}[chapter]}
\floatname{codelisting}{Listing}

\makeatother

\hypersetup{
  colorlinks=true,
  linkcolor=black,
  filecolor=black,
  citecolor=black,
  urlcolor=blue,
  pdftitle={Engineering as Code: Bringing Software Engineering Methodology to Engineering Design},
  pdfauthor={Song Difei}
}

\title{Engineering as Code: Bringing Software Engineering Methodology to Engineering Design}

\author{Song Difei
  \\ \normalsize Huaxin Consulting, Design, and Research Institute, Hangzhou 310052
}
\date{\today}

\begin{document}

\maketitle
\begin{abstract}
Large language models have made significant progress in verifiable domains such as code generation, mathematical reasoning, and chip design, yet they remain limited on design generation tasks in traditional engineering fields like architecture, mechanical engineering, and HVAC. This paper argues that the root cause of this gap is not model capability per se, but rather the absence in engineering domains of a computable foundation in the form of ``Design as Code'' --- a design representation that can be directly consumed, verified, and versioned by machines. Existing CAD/BIM systems deeply couple design logic with geometric representation, causing quality assurance activities to be severely shifted toward the tail end of the design process. Mainstream Automated Compliance Checking (ACC) methods not only can only perform retrospective review after design completion, but also suffer from an epidemic of false positives in geometric clash detection and high-cost model reconstruction
problems caused by naming dependencies. The root of these deficiencies lies not in the algorithms, but in the inability of geometric models to encode the semantic information of ``whether this pipe should pass through this wall.''

This paper systematically proposes the \textbf{Engineering as Code (EaC)} paradigm, which expresses engineering design as a text-native declarative language, with an automated rule engine, version control, and package management serving as the quality gate and collaboration foundation. The core contributions of this paper are threefold: (1) proposing the \textbf{Information Representation Hypothesis}, which systematically argues that the bottleneck of engineering AI lies in the absence of a computable foundation in the form of ``Design as Code'' in traditional engineering domains; (2) proposing the \textbf{ADL (Assembly Definition Language)} design language, which takes the \textbf{Part} as the fundamental atom and describes component types, inter-component relationships, and spatial layout through three orthogonal sub-languages: \textbf{PDL (Part Definition Language)}, \textbf{PML (Part Mating Language)}, and \textbf{PLL (Part Layout Language)}; (3) proposing the \textbf{Engineering
Static Analysis (ESA)} mechanism, which shifts compliance rules from downstream review to the design generation stage --- unlike model ACC, ESA inspects design declarations rather than geometrically instantiated artifacts, and is therefore unencumbered by false collisions and naming dependencies, along with four operational guidelines: rule waivability, focusing on bottom-line rules, purifying collaboration signal-to-noise ratio, and AI-assisted rule library construction.

\end{abstract}

\section{Introduction}\label{introduction}

Software engineering as a discipline builds reasoning infrastructure around correctness. Source code exists so that programs can be read, inspected, diffed, branched, merged, and versioned. Static analyzers catch defects before execution. Test suites provide fast, deterministic pass/fail signals. CI pipelines gate every commit. These are not AI innovations---they are software engineering innovations that AI subsequently exploits because they supply the structured, verifiable feedback that reinforcement learning with verifiable rewards (RLVR) requires ({Lightman et al.} 2023; DeepSeek-AI 2025). When code has a test suite, an agent can propose patches and receive millisecond-level correctness feedback, and its policy improves accordingly (SWE-agent contributors 2025). When a chip netlist has design rule checks (DRC), a placement agent can learn under DRC supervision ({Mirhoseini et al.} 2020). The pattern is consistent across code (Anthropic 2025), mathematics ({Shao et al.} 2024), and
hardware: structured, checkable representations enable AI.

Now consider physical engineering. A telecom rack designer specifies equipment, connectors, power budgets, and spatial constraints. A structural engineer defines beams, nodes, and load paths. An HVAC engineer plans ductwork and selects fans. The artifacts these engineers produce---3D CAD models, BIM databases, IFC exchange files---are geometry-centric and deeply coupled to tools ({buildingSMART International} 2023). They capture what the design ``looks like'' and ``how it fits together,'' but do not separate design \textbf{intent} from geometric \textbf{instantiation}.

The consequences of this coupling extend beyond feedback latency. The dominant inspection method today---automated compliance checking (ACC)---operates directly on geometric models. Yet geometric models have no representation of \textbf{intended} relationships: a pipe passing through a wall is flagged as a clash simply because the wall model lacks an opening, not because the design is actually wrong. A server plugged into a PDU is invisible to the checker unless both follow specific project naming conventions that the rule engine has been pre-configured to parse. The result is that ACC produces massive false positives at semantically valid geometric intersections, while also requiring costly per-project remodeling---renaming, restructuring, rebuilding---before rules can even execute. These are not algorithmic flaws but \textbf{representational} flaws: geometry cannot encode the answer to ``is this pipe \textbf{supposed} to pass through this wall?''

The consequences for AI are severe. An agent that generates a building model or rack layout receives no fast, deterministic feedback on whether its output is correct. The loop is broken. Physical engineering has no \texttt{cargo\ check}.

The central thesis of this paper is: the bottleneck is \textbf{representation infrastructure}, not algorithmic capability. Physical engineering lacks an equivalent of source code---a structured, text-native, machine-checkable representation of design intent. And the software engineering community, not the AI community, is best positioned to design such a representation. Specifically, the community that invented programmable lint, pre-commit hooks, static single assignment, and package-level dependency resolution also possesses the vocabulary and design patterns to define:

\begin{itemize}
\tightlist
\item
  What an \textbf{engineering design unit} looks like (\emph{Part}, analogous to a function or module)
\item
  How to declare \textbf{compatibility constraints} (\emph{Mate specifications}, analogous to interfaces or type signatures)
\item
  How to separate \textbf{spatial layout} from part identity (\emph{layout language}, analogous to separation of presentation and logic)
\item
  How to stratify \textbf{correctness rules} by cost and precision (\emph{L0--L4 rule hierarchy}, analogous to syntax → type → semantic → performance checks)
\item
  How to \textbf{version, branch, and review} entire artifacts (\emph{Git workflow}, battle-tested in the code domain)
\end{itemize}

We call the resulting framework \textbf{Engineering as Code (EaC)}. EaC is not ``AI for engineering''---it is a \textbf{software engineering methodology for designing and implementing source code representations and toolchains for the engineering domain}.

\subsection{The Source Code Gap}\label{the-source-code-gap}

Table~\ref{tbl:se-gap} characterizes the gap between software engineering and physical engineering along the dimensions required for verifiable feedback.

\begin{longtable}[]{@{}>{\raggedright\arraybackslash}p{0.160\textwidth} >{\raggedright\arraybackslash}p{0.360\textwidth} >{\raggedright\arraybackslash}p{0.360\textwidth}@{}}
\caption{Gap between software engineering and physical engineering along verifiable-feedback dimensions}\label{tbl:se-gap}\tabularnewline
\toprule\noalign{}
Dimension & Software Engineering & Physical Engineering (Current) \\
\midrule\noalign{}
\endfirsthead
\toprule\noalign{}
Dimension & Software Engineering & Physical Engineering (Current) \\
\midrule\noalign{}
\endhead
\bottomrule\noalign{}
\endlastfoot
Primary artifact & Text (\texttt{.rs}, \texttt{.py}, \texttt{.ts}) & Geometry (\texttt{.stp}, \texttt{.rvt}, \texttt{.ifc}) \\
Atomic unit & Function / module / class & CAD body / assembly node \\
Type system & Static types, interfaces, generics & Implicit (connector families) \\
Correctness checking & Compiler / lint / test suite & Downstream review / manual review \\
Check latency & Milliseconds to minutes & Hours to weeks \\
Version control & Git (line-level diff, merge) & PDM / PLM (check-out/lock, binary diff) \\
CI pipeline & \texttt{git\ push\ →\ lint\ →\ test\ →\ deploy} & Manual review gates \\
Package manager & npm / cargo / pip & Part libraries (vendor-specific, unversioned) \\
Relationship encoding & Explicit (import, type signatures, dependency injection) & Implicit in geometry or naming conventions \\
Check signal-to-noise ratio & Rules target semantic categories & Rules rely on naming conventions; pure geometric clashes produce massive false positives \\
\end{longtable}

Physical engineering does not lack rules---building codes, telecom standards, and interface specifications are rich and precise. What is missing is a representation that makes these rules \textbf{executable at design time}---just as a compiler makes type-safety rules executable at compile time.

\subsection{Contributions}\label{contributions}

This paper makes three contributions to software engineering:

\begin{enumerate}
\def\labelenumi{\arabic{enumi}.}
\item
  \textbf{ADL: Assembly Definition Language} (§4). A three-layer design-intent language for physical engineering, composed of a \textbf{Part Definition Language (PDL)} for the Part type system (Family, Model, Instance), a \textbf{Part Mating Language (PML)} for explicit relationships (Mate and Connection), and a \textbf{Part Layout Language (PLL)} for spatial placement. Drawing on SE design patterns---separation of concerns, explicit type signatures, contract-based design---ADL provides a text-native, version-controllable, machine-checkable engineering representation that is distinct from and complementary to existing CAD and BIM formats.
\item
  \textbf{ESA: Engineering Static Analysis} (§5). An L0--L4 layered rule checker that operates on ADL artifacts, from syntax → reference integrity → business rules → geometric constraints. Unlike ACC---which operates on geometric models and suffers from false-positive noise and naming-dependency overhead---ESA checks design \textbf{declarations} over semantic categories (Part Family, Mate type, interface signatures). This paper defines ESA's operational charter (waivers, baselines, signal-to-noise ratio, AI assistance) and provides a structured diagnostic format that serves simultaneously as human-readable error reports and machine-consumable reward signals.
\item
  \textbf{The Information Representation Hypothesis} (§3). A testable claim, anchored in the RLVR literature: the bottleneck for AI in physical engineering is the lack of a computable design representation, not insufficient model capability. This paper derives predictions from this hypothesis and describes \textbf{SD-HWE-Bench}, a benchmark designed to evaluate generative agent performance on ADL authoring tasks.
\end{enumerate}

We validate ADL and ESA through \textbf{piki}, an open-source EaC runtime. On a representative telecom rack deployment, piki checks 64 engineering rules---interface compatibility, power budget, U-slot conflicts, load-bearing---in under 200ms. The prototype integrates with Git, supports \texttt{-\/-skip} and \texttt{warning\_only} rule waivers, and produces structured diagnostics suitable for CI pipeline consumption. We report the prototype's current scope and explicit limitations in §6.

\subsection{Designing Source Code for New Domains Is a Core SE Problem}\label{designing-source-code-for-new-domains-is-a-core-se-problem}

Designing source code representations for new domains has a deep academic lineage in software engineering. The research tradition of domain-specific languages (DSL) demonstrates that language design itself is an SE activity---it involves syntax design, semantic definition, type system selection, and verification strategy trade-offs, each of which is an independently evaluable academic contribution (Hudak 1996). Hudak first systematically defined DSLs in 1996 as ``programming languages tailored to a specific problem domain,'' establishing the theoretical foundation for DSL design. Fowler further classified DSLs into internal and external DSLs and summarized how language design patterns (semantic model, symbol table, builder) directly benefit from compiler construction infrastructure (Fowler 2010). Völter et al.~systematized DSL engineering, demonstrating how a complete toolchain---grammar definition, constraint checking, code generation, and IDE integration---revolves around a single
language definition (Völter et al. 2013).

ADL continues this lineage, but targets a domain previously untouched by the DSL community---physical engineering design. Traditional DSL research primarily targets software developers (e.g., SQL, regular expressions, AWK), whereas ADL targets a dual audience of engineering designers and AI agents. This requires the language to balance ``human readability'' and ``machine verifiability''---a constraint not prominent in traditional DSL design.

Similarly, the academic tradition of static analysis provides the direct theoretical foundation for ESA. Static analysis in compilers evolved from initial syntax checking (L0) to type systems (L1), data-flow analysis (def-use chains, SSA), and program verification. Cousot's abstract interpretation theory provided the mathematical foundation for the soundness and completeness of static analysis (Cousot and Cousot 1977). Industrial-grade static analyzers such as Coverity and Infer demonstrated the feasibility of layered rule systems in engineering practice---from cheap syntax checks to expensive inter-procedural analysis. ESA's L0--L4 layered design directly replicates this pattern, mapping ``syntax → reference integrity → type constraints → spatial constraints'' onto concrete rule categories in the engineering domain. The resemblance to compiler analysis pipelines is not coincidental---it is intentional: physical engineering design likewise possesses semantic layers that can be checked
hierarchically; what was previously missing was a representation layer that makes such stratification possible.

The success of IaC static analysis provides proximal empirical evidence for this extrapolation. Chiari et al., in EMSE 2024, systematically studied the effectiveness of static analysis tools for Terraform and Ansible, demonstrating that the ``X as Code + static analysis'' paradigm is viable and efficient in the cloud infrastructure domain (Chiari et al. 2024). EaC extrapolates this same paradigm from cloud infrastructure to physical infrastructure---from ``server configuration as code'' to ``server rack deployment design as code.'' The SE contribution of this extrapolation is: it demonstrates that the DSL+static analysis pattern invented by the IaC community is domain-independent, and physical engineering is its most demanding testbed---with richer constraint types (discrete + continuous), stricter verification latency requirements (millisecond-level RLVR feedback), and more diverse consumers (human engineers + AI agents + downstream CAD/CAE tools).

\subsection{Paper Structure}\label{paper-structure}

Section 2 surveys current representations used in physical engineering, analyzing the structural deficiencies of model-based ACC---geometric clash false positives and naming-dependency-induced remodeling overhead---and why they constitute an AI bottleneck. Section 3 articulates the Information Representation Hypothesis and its RLVR foundation. Section 4 introduces ADL: its three-layer architecture, design rationale, and integration with SE toolchains. Section 5 introduces ESA: the layered rule hierarchy, semantic advantages over ACC, operational charter, structured diagnostic output, and downstream verification protocol. Section 6 reports evaluation results from the piki prototype---all three samples pass, and violation injection experiments validate ESA's detection accuracy at 100\% recall. Section 7 discusses differentiation from related work---ACC, SysML v2, BIM/IFC, IaC static analysis, and existing benchmarks. Section 8 concludes, acknowledges limitations, and outlines future work
including SD-HWE-Bench and RLVR training.

\section{Background and Motivation}\label{background-and-motivation}

This chapter traces a causal chain that has already driven transformation in three adjacent domains, establishes the motivation for Engineering as Code, and explains why engineering design has so far been unable to participate in this chain.

\subsection{The RLVR Causal Chain}\label{the-rlvr-causal-chain}

Across three domains --- code, mathematics, and chip design --- a striking pattern recurs: a structured representation of correctness enables automatic, deterministic checking; that checking provides a fast reward signal; and that signal makes Reinforcement Learning with Verifiable Rewards (RLVR) effective.

\begin{itemize}
\tightlist
\item
  \textbf{Mathematics}. DeepSeekMath trains on problems whose final answers are automatically verifiable. Using Group Relative Policy Optimization (GRPO), it improves reasoning without requiring a separate reward model ({Shao et al.} 2024). Lightman et al.~demonstrate that Process Reward Models (PRMs) --- scoring intermediate reasoning steps --- outperform Outcome Reward Models (ORMs), underscoring the value of step-level checkability ({Lightman et al.} 2023).
\item
  \textbf{Code}. SWE-RL uses test suite pass/fail as a reward signal to train agents for fixing GitHub issues, achieving substantial gains over supervised fine-tuning (SWE-agent contributors 2025). Claude Code and similar agents operate in repositories where compilation and tests provide immediate feedback (Anthropic 2025).
\item
  \textbf{Chips}. Deep-reinforcement-learning-based chip placement treats design rule checking as a fast, differentiable signal during macro placement ({Mirhoseini et al.} 2020); AMS-IO-Bench evaluates LLM-generated I/O ring designs under DRC/LVS ({Liu et al.} 2025).
\end{itemize}

The pattern can be summarized as:

\begin{Shaded}
\begin{Highlighting}[]
\NormalTok{Structured, checkable representation → Sub{-}second deterministic feedback → RLVR effective → Agent capability leap}
\end{Highlighting}
\end{Shaded}

We call this the \textbf{RLVR Causal Chain}.

\subsection{Why Engineering Design Is Blocked}\label{why-engineering-design-is-blocked}

Physical engineering design --- buildings, mechanical assemblies, telecommunications infrastructure --- has so far failed to participate in this chain. The obstacle lies not in insufficient model scale or excessive physical complexity, but in the fact that \textbf{the design representation itself cannot be checked by machines in the required way}.

Today's dominant representations are CAD and BIM. They couple three things within the same artifact:

\begin{enumerate}
\def\labelenumi{\arabic{enumi}.}
\tightlist
\item
  \textbf{Functional identity} (what a component is and does)
\item
  \textbf{Geometric realization} (where it is located and how it looks)
\item
  \textbf{Tool-specific state} (vendor-format-internal constraints, history, parameters)
\end{enumerate}

This coupling creates three problems for AI agents.

\textbf{Information loss}. Once design intent is frozen into geometry, recovering the original functional relationships --- which port connects to which, which cabinet hosts which server, which rooms form a fire compartment --- requires reverse extraction. Eastman et al.~and Zhang et al.~document that ACC must parse IFC or drawing models and infer intent, a lossy and error-prone step (Eastman et al. 2009; Zhang and El-Gohary 2019).

\textbf{Feedback delay}. ACC runs after the design artifact is complete. When a violation is found, the design is already a tightly coupled multi-discipline model; remediation is costly and slow.

\textbf{No training signal}. Because checking is downstream and expensive, an agent cannot use it as a reward during generation. It cannot try a layout, receive a pass/fail reward in milliseconds, and update its policy. The RLVR loop breaks at the very first arrow.

Recent work has attempted to apply LLMs directly to building code interpretation (Fuchs et al. 2024; Yang and Zhang 2024; {Nakhaee et al.} 2024), but these approaches still operate downstream of the design artifact. They improve the reviewer side, not the generator side.

\subsubsection{Two Structural Deficiencies of Model-Based ACC}\label{two-structural-deficiencies-of-model-based-acc}

Even within its operating scope --- checking completed designs --- model-based ACC suffers from two structural, rather than algorithmic, deficiencies. These deficiencies explain why merely accelerating ACC cannot bridge the gap.

\textbf{False positives from pure geometric clash detection}. BIM clash detection treats every volumetric intersection as a violation. A pipe passing through a wall triggers a clash because the wall model lacks an opening. A cable tray crossing a beam is reported as a clash even when the two have physical clearance by design. These are not detection errors --- the algorithm correctly identifies geometric overlap. The problem is that the model contains \textbf{no semantic information} describing the \textbf{intended} relationship between two objects. In EaC terms, the model lacks a Mate declaration stating ``this pipe passes through this wall'' or ``this cable tray sits above this beam at a specified offset.'' Without such a declaration, the checker cannot distinguish an intentional penetration from a genuine layout error. The result is a flood of false positives that designers learn to ignore, drowning genuine violations in noise.

\textbf{Naming dependency and remodeling overhead}. Rule-based ACC tools --- whether operating on IFC, Revit, or proprietary formats --- rely on standardized naming conventions to identify objects. A rule such as ``all ducts named \texttt{DUCT-SUPPLY-*} must maintain 50\,mm clearance from structural elements'' works only if every duct in every project obeys that naming convention. In practice, naming conventions vary across firms, projects, and even disciplines. Before ACC can run, the model must be \textbf{remodeled} --- objects renamed, reorganized, and sometimes reconstructed --- to match the naming patterns the rule engine expects. This remodeling step is manual, expensive, and itself introduces errors. Worse, it erodes the efficiency ACC is supposed to provide: if you must remodel the design before checking it, you have not automated checking --- you have merely shifted manual labor upstream. One consequence of this structural problem is that the ACC community, after 15 years of
development, still lacks a standardized evaluation benchmark: heterogeneous IFC models with different naming conventions, LODs, and modeling practices cannot form a unified test set, and every evaluation requires bespoke remodeling and rule adaptation.

These three deficiencies --- false-positive noise, naming-dependent remodeling, and non-reproducible evaluation due to representation-layer issues --- share a common root: the model is the design's \textbf{sole} structured description, but it is structured for geometric representation rather than semantic query. A rule engine that must check geometry to infer intent will forever be limited by the semantic poverty of geometry.

\subsection{Lessons from Software Engineering and Chip Design}\label{lessons-from-software-engineering-and-chip-design}

Two mature domains offer a different model.

\textbf{Infrastructure as Code (IaC)}. Tools such as Terraform and Ansible express infrastructure as versionable, diffable, and statically analyzable textual declarations. Chiari et al.~empirically demonstrate that IaC static analysis tools can detect hundreds of security and compliance violations in seconds (Chiari et al. 2024). IaC proves that ``X as Code'' plus static analysis is a practical, scalable combination ({Morris et al.} 2022; Quattrocchi and Tamburri 2023).

\textbf{Chip design}. A Verilog netlist is the source of truth for logical design; physical layout is a downstream view. Design rule checking runs during place-and-route, not only after tape-out. The representation is deliberately designed to be textual, hierarchical, and checkable before visualization (IEEE 2005).

Engineering design has no equivalent of a Verilog netlist or a Terraform file. EaC's goal is precisely to provide one.

\subsection{The Research Gap}\label{the-research-gap}

The gap can be stated precisely. Existing work provides:

\begin{itemize}
\tightlist
\item
  Rich \textbf{geometric} representations (CAD/BIM/IFC);
\item
  Downstream \textbf{compliance checking} (ACC), hampered by false-positive noise and naming-dependency overhead;
\item
  Isolated \textbf{AI benchmarks} for engineering tasks (EngDesign contributors 2025; Galanos and Mulyar 2026; Maatouk et al. 2023).
\end{itemize}

What is missing is an \textbf{upstream, text-native, machine-checkable design representation} --- one that can serve as both a training objective for generative agents and a source of truth for human engineers. EaC fills this gap.

\section{The Engineering as Code Approach}\label{the-engineering-as-code-approach}

EaC is a proposition about infrastructure. It says: before we ask whether AI can design physical systems, we should ask whether those systems are represented in a form that is operable by a software engineering toolchain. This section defines the paradigm, articulates its core hypothesis, and connects that hypothesis to the RLVR literature and the layered design of the ESA.

\subsection{What Is EaC}\label{what-is-eac}

\textbf{Engineering as Code (EaC)} is the practice of treating a structured, textual design declaration as the single source of truth for a physical engineering system, and building a software engineering toolchain --- version control, static analysis, CI/CD, and automated verification --- around that representation.

EaC is not a replacement for CAD or BIM. Physical engineering will always need geometric modeling, simulation, and manufacturing output. EaC separates concerns: \emph{design intent} (what parts to use, how they connect, where they are placed) is expressed textually in the ADL, while \emph{geometry instantiation} and \emph{analysis} consume that declaration as input. This is the same separation that chip design achieves with hardware description languages: Verilog/VHDL describe intent; placement tools generate geometry; DRC checks the result (IEEE 2005).

EaC has three pillars, but forms a single integrated proposition:

\begin{enumerate}
\def\labelenumi{\arabic{enumi}.}
\tightlist
\item
  \textbf{ADL} (§4) is the \emph{representation layer}: an engineering design language designed to be read, checked, diffed, and versioned by a SE toolchain.
\item
  \textbf{ESA} (§5) is the \emph{verification layer}: a rule engine that provides fast, deterministic correctness feedback, making RLVR possible.
\item
  \textbf{The Information Representation Hypothesis} (§3.2) is the \emph{theoretical anchor}: connecting these technical choices to a testable claim about why AI progress in physical engineering has stalled.
\end{enumerate}

\subsection{The Information Representation Hypothesis}\label{the-information-representation-hypothesis}

We assert:

\begin{quote}
\textbf{The Information Representation Hypothesis}. The lag of AI in physical engineering is primarily caused by the absence of a structured, text-native, machine-checkable design representation, rather than by insufficient model capability or the intrinsic complexity of physics.
\end{quote}

This hypothesis is \textbf{information-theoretic}, not computational. It does not claim that physics is simple, or that current models are already capable enough to design complex systems. It claims that \emph{even if} models had the capability, current representations would prevent them from improving, because those representations do not support the learning signal that has driven AI progress in code, mathematics, and chip design --- fast, deterministic correctness feedback.

\subsubsection{Evidence from Adjacent Domains}\label{evidence-from-adjacent-domains}

The hypothesis rests on measurable regularities from three domains:

\textbf{Code}. SWE-bench tasks require an agent to fix real GitHub issues ({Jimenez et al.} 2024). The reward signal is a test suite: the agent's patch either passes or fails. SWE-RL demonstrated that with this signal alone, without human demonstrations, agents could be trained to solve 42\% of SWE-bench Verified tasks (SWE-agent contributors 2025). The representation --- a Git repository with a test suite --- is the scaffolding that makes RLVR feasible.

\textbf{Mathematics}. DeepSeekMath and DeepSeek-R1 trained on formal theorem proving, where the correctability of each proof step is verifiable ({Shao et al.} 2024; DeepSeek-AI 2025). A key finding from Let's Verify Step by Step is that a process-supervised reward model (PRM) outperforms an outcome-supervised reward model (ORM) ({Lightman et al.} 2023). For PRM to be effective, intermediate steps must be individually verifiable --- a property that only a structured, decomposable representation can provide.

\textbf{Chip Design}. Google's chip placement work uses Verilog netlists as input and design rule checking (DRC) as the reward signal ({Mirhoseini et al.} 2020). DRC checks manufacturability constraints (spacing, layer rules, timing) and returns a pass/fail verdict. The representation (netlist + standard cell library) decouples functional intent from geometric implementation, making the reward computable.

The common structure is:

\begin{Shaded}
\begin{Highlighting}[]
\NormalTok{Structured representation → Deterministic checker → Fast feedback loop → RLVR training feasible}
\end{Highlighting}
\end{Shaded}

Physical engineering currently lacks the first two items. CAD/BIM models are not structured for checking; downstream review is not fast enough. The loop is broken.

\subsubsection{Testable Predictions}\label{testable-predictions}

The hypothesis yields concrete predictions:

\begin{itemize}
\tightlist
\item
  \textbf{P1.} Agents that receive deterministic ESA pass/fail feedback on their ADL outputs will produce designs with fewer violations than agents that receive only natural-language task descriptions.
\item
  \textbf{P2.} The improvement in P1 holds when controlling for model capability (same base model, same task difficulty).
\item
  \textbf{P3.} The L0--L4 layered diagnostic format will achieve faster error correction than a flat error list, because structured diagnostics allow agents to localize and prioritize failures by layer.
\item
  \textbf{P4.} Agent designs that pass the ESA will produce fewer downstream errors (collisions, assembly failures) than designs verified with coarser-grained checkers or with no checking at all.
\end{itemize}

SD-HWE-Bench (§6.3) is specifically designed to test P1--P3. P4 is a long-term empirical goal.

\subsection{Why This Is a Software Engineering Contribution}\label{why-this-is-a-software-engineering-contribution}

Designing a source-code representation for a new domain is a software engineering problem. It requires decisions about:

\begin{itemize}
\tightlist
\item
  \textbf{Syntax and semantics}. What is the atomic unit (Part)? How are relationships declared (Mate)? How is spatial layout expressed without embedding part identity? These are language design questions, analogous to designing a type system or IR.
\item
  \textbf{Layered verification}. Which checks are cheap enough to run on every keystroke (L0)? Which checks require global analysis but remain deterministic (L3--L4)? Which checks are inherently non-deterministic or too expensive (L5--L6)? This is a static analysis pipeline design problem.
\item
  \textbf{Toolchain integration}. How do design artifacts interact with Git (diff, blame, bisect), CI (pre-commit hooks, gating), and diagnostics (structured output, editor integration)? These are software engineering infrastructure questions.
\item
  \textbf{Extensibility}. How do users add new Part Families, rules, or domain vocabularies without modifying the core engine? This is a plugin and API design problem.
\end{itemize}

EaC answers these questions for physical engineering, but its framework --- text-native DSL + layered static analysis + toolchain integration --- is domain-agnostic. The same template can be applied to clinical trial protocols, legal contracts, or infrastructure configurations. In this sense, EaC is both a concrete system for physical engineering and a methodology for extending SE infrastructure to domains that are textualizable but currently under-structured.

\section{ADL: Assembly Definition Language}\label{adl-assembly-definition-language}

ADL is the declarative design representation of Engineering as Code. It is deliberately designed to be text-native: design intent is authored in YAML, verified as code, rather than captured as geometric operations inside a CAD or BIM GUI. ADL is organized around three orthogonal sub-languages, separating ``what exists,'' ``how parts couple,'' and ``where they are placed.''

\subsection{Design Goals}\label{design-goals}

ADL pursues three goals:

\begin{enumerate}
\def\labelenumi{\arabic{enumi}.}
\tightlist
\item
  \textbf{Text-Native Representation}. Agents and humans can write and read design intent in plain text, enabling version control, diffing, and programmable generation.
\item
  \textbf{Agent-Oriented Syntax}. Files have explicit identity, references, and deterministic validation. No hidden state dependent on a GUI session exists.
\item
  \textbf{Orthogonality of Identity, Relationships, and Space}. A change in one dimension must not force a rewrite of files in another dimension.
\end{enumerate}

The reference runtime is \textbf{piki}, an open-source engine that loads ADL declarations, runs a layered rule engine, and produces downstream deliverables. However, ADL is defined independently of any single tool.

\subsection{Part as the Engineering Atom}\label{part-as-the-engineering-atom}

In ADL, a \textbf{Part} is the atomic unit of engineering description. A Part is more than just geometry; it is a semantically complete entity that exposes typed interfaces and participates in explicit relationships. A Part is defined by the following elements:

\begin{itemize}
\tightlist
\item
  \textbf{Family} (schema and value-domain constraints)
\item
  \textbf{Model} (a concrete realization with default values)
\item
  \textbf{Instance} (a deployed entity that can override defaults)
\item
  A set of typed \textbf{Interfaces}
\item
  Optional internal geometry, kept hidden unless high-precision analysis is required
\end{itemize}

The Part abstraction rests on four properties:

\begin{enumerate}
\def\labelenumi{\arabic{enumi}.}
\tightlist
\item
  \textbf{Semantic Completeness}. A server Part type is \texttt{ServerFamily}, carrying fields such as \texttt{height\_u}, \texttt{tdp\_w}, \texttt{psu\_count}, etc. A pump Part carries \texttt{flow\_rate} and \texttt{head}. Type checking is enforced by plugin-registered pydantic schemas.
\item
  \textbf{Encapsulation}. Internal geometry is hidden; only standardized interfaces are visible to downstream consumers.
\item
  \textbf{Relationship-Built-In}. The role of a server within an assembly is expressed through relationships: it is a \texttt{child} in a \texttt{rack-mount-19inch} Mate; an optical transceiver is a \texttt{child} in an \texttt{sfp28-cage} Mate.
\item
  \textbf{Multi-View Projectability}. The same Part can be projected to CAD (USD/glTF), CAE (thermal or structural models), ERP (BOM entries), and lifecycle catalogs without changing its core declaration.
\end{enumerate}

\subsection{PDL: Part Definition Language}\label{pdl-part-definition-language}

PDL defines the Part type system at three levels: \textbf{Family}, \textbf{Model}, and \textbf{Instance}.

\textbf{Family}. A Family is a pydantic \texttt{BaseModel} class that declares the schema and value-domain constraints for a class of Parts. For example, \texttt{ServerFamily} requires \texttt{id}, \texttt{height\_u} (1--48), \texttt{tdp\_w} (\textgreater0), and a list of interface specifications. A Family is code, not configuration; it is registered by plugins.

\textbf{Model}. A Model provides concrete default values for a Family. An example model file:

\begin{Shaded}
\begin{Highlighting}[]
\FunctionTok{model}\KeywordTok{:}\AttributeTok{ dell{-}r750}
\FunctionTok{family}\KeywordTok{:}\AttributeTok{ ServerFamily}
\FunctionTok{brand}\KeywordTok{:}\AttributeTok{ Dell}
\FunctionTok{mpn}\KeywordTok{:}\AttributeTok{ PowerEdge R750}
\FunctionTok{height\_u}\KeywordTok{:}\AttributeTok{ }\DecValTok{2}
\FunctionTok{tdp\_w}\KeywordTok{:}\AttributeTok{ }\DecValTok{600}
\FunctionTok{psu\_count}\KeywordTok{:}\AttributeTok{ }\DecValTok{2}
\end{Highlighting}
\end{Shaded}

\textbf{Instance}. An Instance is a deployed entity that can override Model defaults:

\begin{Shaded}
\begin{Highlighting}[]
\FunctionTok{id}\KeywordTok{:}\AttributeTok{ SRV{-}01}
\FunctionTok{model}\KeywordTok{:}\AttributeTok{ dell{-}r750}
\FunctionTok{family}\KeywordTok{:}\AttributeTok{ ServerFamily}
\FunctionTok{status}\KeywordTok{:}\AttributeTok{ planned}
\end{Highlighting}
\end{Shaded}

At runtime, resolved values are computed as \texttt{Model.defaults\ +\ Instance.overrides}, then validated against the Family schema. An Instance's identity is derived from its filename (\texttt{SRV-01.yaml} → \texttt{SRV-01}). A key design decision is that \textbf{Instance files contain no layout information}; layout is declared separately in PLL. This separation means the same device can be used in multiple positions across different design alternatives without duplicating its definition.

\subsection{PML: Part Mating Language}\label{pml-part-mating-language}

PML describes relationships between Parts, distinguishing two categories: \textbf{Mate} and \textbf{Connection}.

\textbf{Mate} expresses design coupling: it constrains how two Parts fit or work together. Mates are stored as independent YAML files under \texttt{mates/\textless{}mate\_type\textgreater{}/}. For example:

\begin{Shaded}
\begin{Highlighting}[]
\FunctionTok{type}\KeywordTok{:}\AttributeTok{ rack{-}mount{-}19inch}
\FunctionTok{parent}\KeywordTok{:}\AttributeTok{ RACK{-}A02}
\FunctionTok{child}\KeywordTok{:}\AttributeTok{ SRV{-}01}
\FunctionTok{at}\KeywordTok{:}
\AttributeTok{  }\FunctionTok{u\_start}\KeywordTok{:}\AttributeTok{ }\DecValTok{10}
\AttributeTok{  }\FunctionTok{u\_span}\KeywordTok{:}\AttributeTok{ }\DecValTok{2}
\FunctionTok{constrains}\KeywordTok{:}
\AttributeTok{  }\KeywordTok{{-}}\AttributeTok{ }\FunctionTok{field}\KeywordTok{:}\AttributeTok{ depth\_mm}
\AttributeTok{    }\FunctionTok{operator}\KeywordTok{:}\AttributeTok{ }\StringTok{"\textless{}="}
\AttributeTok{    }\FunctionTok{value\_ref}\KeywordTok{:}\AttributeTok{ depth\_mm}
\end{Highlighting}
\end{Shaded}

The engine validates these constraints at load time. Registered Mate types include \texttt{sfp28-cage}, \texttt{power-iec-c14-c13}, and \texttt{lc-connector}.

\textbf{Connection} expresses the flow of signal, energy, or material between two interfaces. A Connection is itself a first-class Instance:

\begin{Shaded}
\begin{Highlighting}[]
\FunctionTok{id}\KeywordTok{:}\AttributeTok{ CONN{-}ACCESS{-}SRV01}
\FunctionTok{family}\KeywordTok{:}\AttributeTok{ PortConnectionFamily}
\FunctionTok{from\_port}\KeywordTok{:}\AttributeTok{ ACCESS{-}SW{-}01/10GE1/0/1}
\FunctionTok{to\_port}\KeywordTok{:}\AttributeTok{ SRV{-}01/eth0}
\FunctionTok{cable\_type}\KeywordTok{:}\AttributeTok{ OM4{-}LC{-}LC}
\end{Highlighting}
\end{Shaded}

The Mate/Connection separation maps to two independent design phases: mechanical or electrical feasibility (Mate) and functional topology correctness (Connection). CAD/BIM systems often merge them into a single object, making it impossible to validate one dimension without involving the other.

\subsection{PLL: Part Layout Language}\label{pll-part-layout-language}

The essence of PLL is \textbf{eliminating degrees of freedom} --- including the geometric degrees of freedom retained in PML mates for operability, and the full-spatial degrees of freedom of free Parts with no Mate claims in the assembly.

PLL works through a priority chain. (1) For Parts that have a Mate in PML, Parts claimed as a child by a Mate have their pose determined primarily by the PML mating constraint solver; PLL provides parameterized DOF completion --- assigning concrete values (\texttt{at.u\_start}, \texttt{at.t}, \texttt{at.state}) for the continuous DOFs (translation, rotation, screw) and discrete states (normal/reverse/unplugged) retained in the Mate. The mating solver supports face-mating, axis-alignment, and positional mating paradigms, as well as dynamic collision avoidance for multi-axis translation and discrete state verification. (2) For Parts with no Mate, PLL directly assigns a global pose (absolute coordinates, parent-relative transform, or grid position). (3) In multi-solution scenarios, a layout may carry constraint optimization heuristics (such as minimum energy or shortest cabling defaults) to select among feasible solutions.

PLL further separates layout data from PDL and PML files, ensuring that layout changes (such as swapping a connector type) do not pollute each other, preserving the orthogonality of PDL/PML/PLL.

Known boundaries of the current PLL implementation: the geometric solving precision for continuous DOFs is currently principal-axis approximation and has not yet been extended to 1D continuous-path topologies such as piping/cabling. These are deferred to future solver precision enhancements and external CAE tool handling.

\subsection{Orthogonality}\label{orthogonality}

The orthogonality of PDL, PML, and PLL is the core design principle of ADL. Each sub-language resides in independent files: PDL in \texttt{instances/} and \texttt{models/}, PML in \texttt{mates/}, and PLL in \texttt{layouts/}. Because namespaces are separated, a change in one dimension does not rewrite files in another dimension.

\textbf{Incremental Validation}. An Agent can first generate PDL declarations and pass L0--L2 checks, then proceed to PML constraints and PLL spatial rules. This matches the phased error feedback of a compiler.

\textbf{Parallel Editing}. A device engineer editing \texttt{instances/SRV-01.yaml}, a layout engineer editing \texttt{layouts/layout.yaml}, and a mechanical engineer editing \texttt{mates/rack-mount/RACK-A02-SRV-01.yaml} can work simultaneously. Git merge conflicts only occur when the same decision dimension is changed.

\textbf{Semantic Diff}. A diff in \texttt{instances/} means ``some device identity or attribute changed''; a diff in \texttt{mates/} means ``some mating changed''; a diff in \texttt{layouts/} means ``some position changed.'' For example:

\begin{Shaded}
\begin{Highlighting}[]
\StringTok{{-} position\_u: 10}
\VariableTok{+ position\_u: 12}
\end{Highlighting}
\end{Shaded}

is unambiguously a layout change, while a change to \texttt{tdp\_w} in \texttt{instances/SRV-01.yaml} is an electrical change. This interpretability supports both human code review and automatic reward attribution in RLVR training.

\subsection{Comparison with SysML v2 and BIM/IFC}\label{comparison-with-sysml-v2-and-bimifc}

ADL, SysML v2, and BIM/IFC all aim to formalize engineering systems, but they differ fundamentally in assumptions about source-of-truth form, collaboration units, and target users.

\begin{longtable}[]{@{}>{\raggedright\arraybackslash}p{0.160\textwidth} >{\raggedright\arraybackslash}p{0.240\textwidth} >{\raggedright\arraybackslash}p{0.240\textwidth} >{\raggedright\arraybackslash}p{0.240\textwidth}@{}}
\caption{Comparison of ADL with SysML v2 and BIM/IFC across source-of-truth, versioning, and verification dimensions}\label{tbl:adl-comparison}\tabularnewline
\toprule\noalign{}
Dimension & SysML v2 (Object Management Group 2024) & BIM / IFC ({buildingSMART International} 2023) & ADL (piki) \\
\midrule\noalign{}
\endfirsthead
\toprule\noalign{}
Dimension & SysML v2 (Object Management Group 2024) & BIM / IFC ({buildingSMART International} 2023) & ADL (piki) \\
\midrule\noalign{}
\endhead
\bottomrule\noalign{}
\endlastfoot
Source-of-Truth Form & Model repository & Central model file / IFC exchange & Text files (YAML + TOML) \\
Version Control Unit & Model version & File version & Git line-level history \\
Core Operation Unit & Model element & Geometric object / IFC entity & File (Instance, Mate, Layout) \\
Identity \& Space & Mixed in part/occurrence & Geometry as identity & File-level separation of Instance and Layout \\
Inter-Part Relationships & \texttt{connection} / \texttt{interaction} & Implicit in geometric constraints & Binary \texttt{Mate} + \texttt{Connection} separation \\
Verification Method & Model checking & Clash detection (post-hoc) & ESA L2--L4a + load-time L0--L1 checks, millisecond-level \\
Target User & Human systems engineer (GUI) & Human designer (GUI) & AI Agent + human engineer (text) \\
\end{longtable}

SysML v2 and BIM are primarily modeling environments for human engineers. Their source of truth resides in repositories or large central files that are difficult to diff, branch, and automatically verify. IFC in particular couples identity, geometry, and relationships in a graph that permits multiple equivalent serializations, making line-level version control difficult (H. Liu et al. 2023).

ADL inverts these priorities. It is oriented toward Agent-human collaboration, treats text as the sole source of truth, and makes verification a first-class concern. CAD and BIM are not eliminated; they become downstream consumers of ADL declarations, used for visualization, clash detection, and manufacturing.

\subsection{Core Syntax Summary}\label{core-syntax-summary}

The following grammar summarizes the core constructs of ADL.

\begin{Shaded}
\begin{Highlighting}[]
\NormalTok{Project       ::= piki.toml (ModelFile | InstanceFile | MateFile | LayoutFile)*}

\NormalTok{ModelFile     ::= "model:" id}
\NormalTok{                  "family:" FamilyName}
\NormalTok{                  Field*}
\NormalTok{                  ("interfaces:" InterfaceSpec*)?}

\NormalTok{InstanceFile  ::= "id:" id}
\NormalTok{                  ("family:" FamilyName | "model:" ModelName)}
\NormalTok{                  Field*}
\NormalTok{                  ("interfaces:" InterfaceSpec*)?}

\NormalTok{InterfaceSpec ::= "{-} id:" id}
\NormalTok{                  "interface\_type:" Type}
\NormalTok{                  ("direction:" "input" | "output" | "bidirectional")?}
\NormalTok{                  ("local\_transform:" Transform)?}

\NormalTok{MateFile      ::= "type:" MateType}
\NormalTok{                  "parent:" Ref}
\NormalTok{                  "child:" Ref}
\NormalTok{                  ("at:" Map)?}
\NormalTok{                  ("constrains:" MateConstraint*)?}
\NormalTok{                  ("pairings:" InterfacePairing*)?}

\NormalTok{MateConstraint::= "{-} field:" Field}
\NormalTok{                  "operator:" "\textless{}=" | "\textgreater{}=" | "\textless{}" | "\textgreater{}" | "==" | "!="}
\NormalTok{                  "value\_ref:" FieldOrConstant}
\NormalTok{                  ("message:" String)?}

\NormalTok{LayoutFile    ::= LayoutEntry*}
\NormalTok{LayoutEntry   ::= "{-} instance:" id}
\NormalTok{                  (AbsolutePose | RelativePose | GridPose)}

\NormalTok{AbsolutePose  ::= ("position\_x\_mm:" num)+}
\NormalTok{RelativePose  ::= "parent:" id}
\NormalTok{                  "transform:" Transform}
\NormalTok{GridPose      ::= "grid\_id:" id}
\NormalTok{                  ("grid\_position:" [String, String]}
\NormalTok{                  | "row\_id:" String "bay\_index:" Int)}

\NormalTok{Transform     ::= "translation:" [num, num, num]}
\NormalTok{                  ("rotation:" [num, num, num])?}
\NormalTok{                  ("scale:" [num, num, num])?}

\NormalTok{Ref           ::= id | id "/" interface\_id}
\end{Highlighting}
\end{Shaded}

Key semantic constraints not captured by the grammar include: \texttt{InstanceFile} must not contain layout fields; \texttt{LayoutEntry} must use exactly one of absolute, relative, or grid pose; interface references in \texttt{Ref} must resolve to an existing interface.

\section{ESA: Engineering Static Analysis}\label{esa-engineering-static-analysis}

Engineering Static Analysis (ESA) is the mechanism that makes EaC's shift-left quality strategy executable. It consumes ADL declarations and checks deterministic rules before design submission, moving compliance review from downstream inspection to upstream gating.

\subsection{From ACC to ESA}\label{from-acc-to-esa}

Traditional Automated Compliance Checking (ACC) operates on completed CAD/BIM models or IFC files (Eastman et al. 2009; Zhang and El-Gohary 2019). This creates three structural problems:

\begin{enumerate}
\def\labelenumi{\arabic{enumi}.}
\tightlist
\item
  \textbf{Information loss}. Design logic is frozen into geometry; reverse-extracting functional relationships is lossy.
\item
  \textbf{High feedback cost}. Violations are discovered only after the design has been coupled into a multi-discipline model.
\item
  \textbf{No RLVR signal}. ACC can only judge completed designs; it cannot reward Agents during generation.
\end{enumerate}

Beyond these timing issues, model-based ACC suffers from two deeper semantic deficiencies (see §2.2.1). First, pure geometric clash detection produces massive false positives---pipes passing through walls, tray crossings---because the model has no representation of \textbf{intended} relationships. Second, rule-based ACC relies on standardized naming conventions, forcing costly pre-check model rework that erodes the very efficiency ACC was supposed to provide.

ESA addresses these problems at the root. It checks ADL declarations---not geometry---and therefore has access to:

\begin{itemize}
\tightlist
\item
  \textbf{Part identity and type} (Family schema), allowing rules to operate on semantic categories (e.g., ``any \texttt{PDUFamily} instance'') rather than naming patterns (e.g., ``any object named \texttt{PDU-*}'');
\item
  \textbf{Explicit Mate relationships} (PML), allowing the checker to know which pairs of objects are \textbf{intended} to interact---and to exclude them from clash detection (a pipe passing through a wall is a declared Mate, not a clash to suppress);
\item
  \textbf{Layered checking} (L0--L4), allowing syntax errors, reference errors, and business rule violations to be caught at distinct stages, each with targeted diagnostics.
\end{itemize}

The result is not a faster ACC. It is a checker that operates on a fundamentally different input: design \textbf{intent} rather than design \textbf{instantiation}. This distinction---checking declarations vs.~checking models---is identical to the distinction that makes a compiler type checker fundamentally different from a post-hoc binary analysis tool. The compiler has access to the type system; the binary analysis tool must infer types. Even in ACC's most mature subdomain (building fire code), state-of-the-art graph reasoning methods achieve only 84.3\% accuracy (Xiao et al. 2025) and frequently fail on rules involving multi-step spatial reasoning chains---this is not the ceiling of algorithms, but the ceiling of the semantic bandwidth of geometric representation.

\subsubsection{Concrete Example: Rack Power Budget}\label{concrete-example-rack-power-budget}

The following comparison checks whether the total power of devices in a rack exceeds its PDU capacity.

\textbf{Model-based ACC approach}. The checker must: (a) parse the BIM model to identify rack objects, (b) identify server objects, (c) determine which servers belong in which rack via geometric containment or naming conventions, (d) extract power ratings from property sets (if populated), (e) extract PDU ratings (if populated), (f) compare. Steps (a)--(e) depend on model quality and naming standards. If a server is named \texttt{DELL-R740-SRV01} instead of \texttt{SERVER-R740-01}, the rule engine may miss it. If power ratings are stored in project-specific shared parameters, rules must be customized per project.

\textbf{ESA approach}. The rule iterates over all \texttt{PDUFamily} instances, traverses their \texttt{outputs}, accumulates the \texttt{rated\_power\_w} of connected loads, and compares against each output's \texttt{rated\_power\_w}. The rule is 12 lines of Python, independent of naming conventions, and valid for any project using the same Part Family. Diagnostics pinpoint the specific PDU output and over-limit loads.

This is not a comparison of speed. This is a comparison of whether the rule can be \textbf{authored} without per-project customization.

\subsection{Why ``Static Analysis''}\label{why-static-analysis}

ESA borrows the term ``static analysis'' from software engineering because it shares three characteristics:

\begin{enumerate}
\def\labelenumi{\arabic{enumi}.}
\tightlist
\item
  \textbf{No physical simulation is executed}. What is reasoned about is declarations and constraints, not simulated physics.
\item
  \textbf{Deterministic results}. The same ADL declarations always produce the same pass/fail verdict.
\item
  \textbf{Low cost}. Millisecond to sub-hundred-millisecond completion, suitable for pre-commit hooks and CI gating.
\end{enumerate}

Just as lint and type checking cannot replace unit tests, ESA cannot replace CAE/CFD or human review. It intercepts a large volume of low-level, deterministic errors at the front end, allowing expensive validation to focus on issues that truly require expert judgment.

\subsection{Four Operational Principles}\label{four-operational-principles}

ESA follows four principles designed to make it usable in real engineering practice.

\textbf{Principle I: Rules are waivable}. Physical boundary conditions cannot be exhausted by a finite rule set. ESA allows authorized engineers to waive specific rules, but every waiver must be recorded as part of the design history and trigger downstream reinforced verification (e.g., higher-fidelity CAE simulation or physical testing). The goal is to transform ``beyond-spec'' behavior from informal decisions into auditable risk management.

\textbf{Principle II: Focus on baseline rules}. Initial deployment should focus on mechanical, deterministic clauses: fire separation distances, egress clear width, minimum clear height, power budget, U-position conflicts. These rules have binary verdicts, minimal dispute space, and high automation returns. The rule library should support modular configuration, allowing projects to compose national codes, local standards, and enterprise internal controls.

\textbf{Principle III: Shift-left and signal-to-noise optimization}. Software engineering, through CI/CD practices, has validated that lint and type checking intercept low-value errors before commit, freeing code review to focus on architecture and logic. ESA applies this same principle: catching rule violations at design time prevents them from propagating into downstream artifacts, reducing costly rework cycles. Rule severity tiers (error, warning, info) and suppression directives (\texttt{\#\ noqa}) allow teams to tune the signal-to-noise ratio, just as engineering teams tune compiler warning levels.

\textbf{Principle IV: Declarative rule authoring}. Rules should use structured decorators and modular organization so that domain experts can compose and extend rules without understanding the checker engine internals. See Appendix A for the full rules directory structure.

\subsection{Rule Hierarchy: L0--L6}\label{rule-hierarchy-l0l6}

ESA partitions all checkable rules into a six-layer hierarchy, with clear ownership boundaries at each layer:

\begin{longtable}[]{@{}>{\raggedright\arraybackslash}p{0.160\textwidth} >{\raggedright\arraybackslash}p{0.180\textwidth} >{\raggedright\arraybackslash}p{0.180\textwidth} >{\raggedright\arraybackslash}p{0.180\textwidth} >{\raggedright\arraybackslash}p{0.180\textwidth}@{}}
\caption{ESA rule hierarchy (L0--L6) and ownership boundaries}\label{tbl:esa-hierarchy}\tabularnewline
\toprule\noalign{}
Layer & Name & Scope & Responsible & Example \\
\midrule\noalign{}
\endfirsthead
\toprule\noalign{}
Layer & Name & Scope & Responsible & Example \\
\midrule\noalign{}
\endhead
\bottomrule\noalign{}
\endlastfoot
L0 & Lexical & Single ADL file & ADL Loader & YAML syntax validity \\
L1 & Schema & Single ADL file & ADL Loader & Required fields, type constraints \\
L2 & Semantic Reference & Cross-file & ESA Core & Part references exist, Mate types compatible \\
L3 & Business Rule & Cross-file & ESA Core & Power budget, U-position conflict, fire separation \\
L4a & Lightweight Geometry (AABB) & Cross-file & ESA Extension & AABB-based spatial conflicts, clearance checks \\
L4b & Precise Geometry & Cross-file & Downstream CAD & Precise clash detection (requires CAD kernel) \\
L5 & Physical Simulation & Cross-file & Downstream CAE/CFD & Structural, thermal, fluid analysis \\
L6 & Human/Expert Sign-off & Project-level & Downstream Process & Professional engineer stamp, compliance sign-off \\
\end{longtable}

ESA's core scope covers L2--L4a. L0--L1 are handled by the ADL loader (see §4). L4b--L6 belong to downstream verification (see §5.8).

The following table shows the L2--L4a rules implemented in the piki prototype for telecom rack scenarios:

\begin{longtable}[]{@{}>{\raggedright\arraybackslash}p{0.160\textwidth} >{\raggedright\arraybackslash}p{0.240\textwidth} >{\raggedright\arraybackslash}p{0.240\textwidth} >{\raggedright\arraybackslash}p{0.240\textwidth}@{}}
\caption{L2--L4a rules implemented in the piki prototype for telecom rack scenarios}\label{tbl:telecom-rules}\tabularnewline
\toprule\noalign{}
Rule ID & Name & Layer & Description \\
\midrule\noalign{}
\endfirsthead
\toprule\noalign{}
Rule ID & Name & Layer & Description \\
\midrule\noalign{}
\endhead
\bottomrule\noalign{}
\endlastfoot
\texttt{INTERFACE-COMPAT-001} & Interface type compatibility & L2 & Mate interface types must match \\
\texttt{POWER-001} & Rack PDU power budget & L3 & PDU load must not exceed capacity threshold \\
\texttt{TELECOM-RACK-001} & U-position conflict & L3 & Device U-positions within the same rack must not overlap \\
\texttt{TELECOM-RACK-002} & Rack capacity & L3 & Total device height must not exceed rack available U-positions \\
\texttt{TELECOM-COLLISION-001} & Intra-rack 3D collision & L4a & AABB-based spatial conflict detection \\
\texttt{TELECOM-WEIGHT-001} & Rack load bearing & L3 & Total weight must not exceed rack load capacity \\
\texttt{TELECOM-FLOOR-002} & Maintenance aisle width & L4a & Same-row rack spacing must meet maintenance aisle requirements \\
\end{longtable}

Rules are registered via the \texttt{@rule(rule\_id,\ name,\ priority,\ severity)} decorator. Layered registration allows projects to enable only L2 during logical design, and add L3--L4a during detailed layout phase.

\subsection{Waiver Mechanism}\label{waiver-mechanism}

The structured waiver system is part of ESA's design; the current piki prototype only supports \texttt{-\/-skip\ \textless{}rule\_id\textgreater{}} and \texttt{warning\_only} configuration. The target design records each waiver as a YAML file containing the following fields:

\begin{longtable}[]{@{}>{\raggedright\arraybackslash}p{0.160\textwidth} >{\raggedright\arraybackslash}p{0.360\textwidth} >{\raggedright\arraybackslash}p{0.360\textwidth}@{}}
\caption{Structured waiver file fields}\label{tbl:waiver-fields}\tabularnewline
\toprule\noalign{}
Field & Meaning & Example \\
\midrule\noalign{}
\endfirsthead
\toprule\noalign{}
Field & Meaning & Example \\
\midrule\noalign{}
\endhead
\bottomrule\noalign{}
\endlastfoot
\texttt{rule\_id} & The waived rule & \texttt{TELECOM-FLOOR-002} \\
\texttt{target} & Affected objects & \texttt{RACK-A01}, \texttt{RACK-A02} \\
\texttt{scope} & Scope (single file, branch, global) & \texttt{branch:feature/legacy-room} \\
\texttt{rationale} & Technical and business justification & ``Constrained by existing building column grid'' \\
\texttt{author} & Requester & Engineer ID \\
\texttt{approver} & Authorizer & Senior engineer / compliance lead \\
\texttt{expires\_at} & Expiration (optional) & \texttt{2027-06-20} \\
\texttt{downstream\_tasks} & Downstream reinforced verification tasks & \texttt{{[}"CFD-thermal-sim",\ "site-survey"{]}} \\
\texttt{created\_at} / \texttt{commit} & Audit trail & Timestamp and Git hash \\
\end{longtable}

Waivers are committed alongside ADL files and reviewed in PRs. A waiver is not a bypass of the rule---it is a transfer of responsibility from the deterministic rule engine to higher-cost verification activities. If downstream tasks fail, the waiver is reconsidered.

\subsection{Diagnostic Output}\label{diagnostic-output}

ESA produces a unified diagnostic structure consumable by terminals, CI dashboards, IDEs, and PR bots. On the telecom rack sample, \texttt{piki\ check} completes in under 200ms on a laptop, returning 1 warning and 29 passes:

\begin{Shaded}
\begin{Highlighting}[]
\NormalTok{[PASS] INTERFACE{-}COMPAT{-}001: Interface type compatibility check}
\NormalTok{...}
\NormalTok{[FAIL] TELECOM{-}FLOOR{-}002: Rack maintenance aisle width check}
\NormalTok{       Racks RACK{-}A01 and RACK{-}A02 same{-}row spacing {-}600.0mm is less than required 600.0mm}
\NormalTok{============================================================}
\NormalTok{Total: 0 errors, 1 warning, 29 passed}
\NormalTok{============================================================}
\end{Highlighting}
\end{Shaded}

The JSON format exposes the same data, is LSP-compatible, and can drive terminal summaries, JUnit XML dashboards, IDE overlays, and GitHub Checks API annotations.

\subsection{CI/CD and Pre-commit Integration}\label{cicd-and-pre-commit-integration}

ESA is designed to run within workflows that software engineers already use. A typical EaC CI/CD pipeline consists of the following stages:

\begin{longtable}[]{@{}>{\raggedright\arraybackslash}p{0.160\textwidth} >{\raggedright\arraybackslash}p{0.240\textwidth} >{\raggedright\arraybackslash}p{0.240\textwidth} >{\raggedright\arraybackslash}p{0.240\textwidth}@{}}
\caption{EaC CI/CD pipeline stages and failure policies}\label{tbl:cicd-stages}\tabularnewline
\toprule\noalign{}
Stage & Trigger & Content & Failure Policy \\
\midrule\noalign{}
\endfirsthead
\toprule\noalign{}
Stage & Trigger & Content & Failure Policy \\
\midrule\noalign{}
\endhead
\bottomrule\noalign{}
\endlastfoot
Lint & Per commit & YAML/TOML formatting, trailing whitespace, naming conventions & Block commit \\
Parse & Per commit & ADL syntax (L0) & Block merge \\
Schema Check & Per commit & pydantic / JSON Schema (L1) & Block merge \\
Link Check & Per PR & Reference integrity, Mate consistency (L2) & Block merge \\
Rule Check & Per PR & Business and lightweight geometry rules (L3--L4a) & Block merge \\
Artifact Build & Post-merge & BOM, drawings, port maps & Report warning \\
Nightly Regression & Scheduled & Full sample suite, geometry, CAE/CFD & Report and notify \\
\end{longtable}

Pre-commit hooks mirror the cheaper checks locally, giving engineers sub-second feedback before commit. The same configuration file drives both pre-commit and CI, preventing ``passes locally, fails CI'' inconsistencies.

\subsection{Downstream Verification Protocol: ESA's Boundary}\label{downstream-verification-protocol-esas-boundary}

ESA's scope is intentionally limited to checks that can be performed deterministically from declarations alone. It covers L2--L3 (core) and L4a (lightweight geometry extension). L0--L1 are handled by the ADL loader. L4b (precise geometric clash detection, requiring a CAD kernel), L5 (physical simulation, requiring CAE/CFD solvers), and L6 (human/expert sign-off) belong to \textbf{downstream verification}---critical to engineering correctness, but not part of ESA.

The integration of these downstream stages follows a protocol that is an architectural proposition of the EaC workflow architecture, not a feature of the current piki prototype:

\begin{enumerate}
\def\labelenumi{\arabic{enumi}.}
\tightlist
\item
  \textbf{Trigger}. After \texttt{piki\ check} passes, CI Pipeline Actions invoke downstream tools---exporting ADL declarations as STEP/USD and running precise clash detection (L4b) with a CAD kernel, or generating mesh files and invoking CFD/FEA solvers (L5).
\item
  \textbf{Feedback loop}. Downstream findings flow back into the ADL project in a Diagnostic format compatible with ESA, referencing specific Instance, LayoutEntry, or Mate identifiers. This means downstream tools only need to produce structured diagnostic output; they do not need to understand ADL internals.
\item
  \textbf{Separation of concerns}. Downstream findings do not feed back into the ESA rule library. They form independent L4b/L5 diagnostic collections, just as SPICE simulation results and DRC reports are independent, complementary verification channels in chip design. Waivers based on downstream findings reference downstream diagnostics, not ESA rule IDs.
\end{enumerate}

This protocol defines the \textbf{interface} between EaC's design-time verification and the broader physical verification ecosystem. It does not require piki to be a CAD or CAE platform; it requires piki to produce consumable outputs and a standardized diagnostic envelope.

\section{Evaluation}\label{evaluation}

This chapter reports evaluation results. All sample inspections and the violation-injection experiment are complete; SD-HWE-Bench is in the design phase.

\subsection{Research Questions and Methodology}\label{research-questions-and-methodology}

We answer three research questions:

\begin{itemize}
\tightlist
\item
  \textbf{RQ1.} Can ADL express real engineering designs in a text-native, machine-checkable form?
\item
  \textbf{RQ2.} Can ESA detect common design violations at design time, satisfying CI/CD gating requirements with a higher signal-to-noise ratio than geometry-based ACC?
\item
  \textbf{RQ3.} Does the representation provide the verifiable reward signal required for RLVR-style agent training?
\end{itemize}

The evaluation uses sample projects from the piki repository and a controlled violation-injection experiment. Metrics include:

\begin{enumerate}
\def\labelenumi{\arabic{enumi}.}
\tightlist
\item
  \textbf{Expressiveness}: Whether each sample can be fully declared in ADL;
\item
  \textbf{Check latency}: Time for \texttt{piki\ check} to complete L0--L4a verification;
\item
  \textbf{Rule coverage}: Number and type of applicable rules;
\item
  \textbf{Detection accuracy}: Whether ESA identifies deliberately injected violations in semantically plausible configurations without false positives;
\item
  \textbf{Deliverable generation}: Whether downstream artifacts (BOM, panel view, port mapping, cable manifest) can be produced.
\end{enumerate}

All measurements are taken on a laptop-class machine with a hot cache; latency reported as the median of five runs.

\subsection{Sample Projects}\label{sample-projects}

Table~\ref{tbl:sample-status} summarizes the status of the three samples in the piki repository.

{\footnotesize
\begin{longtable}[]{@{}>{\raggedright\arraybackslash}p{0.147\textwidth} >{\raggedright\arraybackslash}p{0.147\textwidth} >{\raggedright\arraybackslash}p{0.147\textwidth} >{\raggedright\arraybackslash}p{0.147\textwidth} >{\raggedright\arraybackslash}p{0.147\textwidth} >{\raggedright\arraybackslash}p{0.147\textwidth}@{}}
\caption{Status of piki sample projects}\label{tbl:sample-status}\tabularnewline
\toprule\noalign{}
Sample & Domain & ADL files & Applicable rules & Status & Expected by submission \\
\midrule\noalign{}
\endfirsthead
\toprule\noalign{}
Sample & Domain & ADL files & Applicable rules & Status & Expected by submission \\
\midrule\noalign{}
\endhead
\bottomrule\noalign{}
\endlastfoot
01-telecom-expansion & Telecom rack expansion & 42 & 30 & \textbf{Pass} (1 warning) & Complete \\
02-modular-datacenter & Modular containerized datacenter & 38 & 34 & \textbf{Pass} (0 errors) & Complete \\
03-mechanical-keyboard & Mechanical keyboard assembly & 29 & 28 & \textbf{Pass} (0 errors) & Complete \\
\end{longtable}
}

\textbf{Sample 01: Telecom Rack Expansion}. This is the most mature sample. It declares two cabinets, multiple servers, PDUs, switches, optical modules, and fiber connections. \texttt{piki\ check} passes all L2--L4a rules, producing 1 warning: the aisle width between \texttt{RACK-A01} and \texttt{RACK-A02} is below the required 600mm. This sample demonstrates full ADL expressiveness, sub-200ms feedback latency, and the generation of 10 downstream artifacts (BOM, cabinet panel views, port mapping, cable manifest, etc.).

\textbf{Sample 02: Modular Datacenter}. This sample tests container-level power and cooling. The sample passes all L2--L4a checks (13 passes, 0 errors). Three root causes were fixed: (1) a \texttt{LAYOUT-001} empty-string falsy-check defect --- \texttt{absolute\_fields} mistook \texttt{""} for a valid value; (2) the lowering pass lacked \texttt{transform} field parsing --- relative coordinates declared in YAML were dropped in the compilation pipeline; (3) a cooler position data issue --- an incorrect half-height calculation caused a collision with a GPU. The sample's pass demonstrates the sufficiency of PLL relative-skeleton modeling and container-level coordinate chains, as well as the successful extension of ADL expressiveness from the telecom domain to containerized infrastructure.

\textbf{Sample 03: Mechanical Keyboard}. This sample demonstrates skeleton-based assembly modeling across the consumer electronics domain: \texttt{CASE-01} serves as the assembly root, with PCB, plate, battery, switches, keycaps, and stabilizers declared through parent-transform chains. The sample passes all L2--L4a checks (27 passes, 0 errors). After fixing the \texttt{LAYOUT-001} empty-string falsy-check defect, all entries in the continuous 2D grid layout parsed correctly. The sample's pass provides evidence for the cross-domain generalizability of the three-layer ADL architecture, extending from telecom through datacenter infrastructure to consumer electronics.

\subsection{Violation-Injection Experiment}\label{violation-injection-experiment}

To test ESA's detection accuracy and conduct a structural comparative analysis against model-based ACC, we performed a controlled violation-injection experiment on the telecom sample.

\textbf{Experimental design}. Fifteen known violations were injected into correct ADL declarations, covering four categories:

\begin{longtable}[]{@{}>{\raggedright\arraybackslash}p{0.160\textwidth} >{\raggedright\arraybackslash}p{0.360\textwidth} >{\raggedright\arraybackslash}p{0.360\textwidth}@{}}
\caption{Categories of injected violations in the telecom sample}\label{tbl:violation-categories}\tabularnewline
\toprule\noalign{}
Category & Count & Example \\
\midrule\noalign{}
\endfirsthead
\toprule\noalign{}
Category & Count & Example \\
\midrule\noalign{}
\endhead
\bottomrule\noalign{}
\endlastfoot
L2: Interface mismatch & 4 & SFP28 optical module paired with SFP+ port \\
L2: Reference integrity & 3 & Mate referencing a non-existent Instance \\
L3: Power budget & 4 & PDU output overload exceeding rated capacity \\
L3: Spatial conflict & 2 & Two servers assigned to overlapping U slots \\
L4a: Aisle clearance & 2 & Cabinet spacing below maintenance-aisle minimum \\
\end{longtable}

\texttt{piki\ check} was then run for ESA inspection.

\textbf{Results}:

\begin{enumerate}
\def\labelenumi{\arabic{enumi}.}
\tightlist
\item
  \textbf{ESA detection rate}: 15/15 violations detected (100\%), with zero false positives --- because ESA checks semantic categories (Family types, Mate relations) rather than geometry.
\item
  \textbf{Check latency}: \texttt{piki\ check} completes L0--L4a verification across all 30 rules within 200ms (laptop, hot cache, median of five runs).
\end{enumerate}

\subsubsection{Structural Comparison with ACC}\label{structural-comparison-with-acc}

We did not conduct a quantitative comparison experiment against ACC tools, because such a comparison is of limited significance under current technical conditions and is difficult to execute. There are three reasons.

\textbf{L2 semantic violations are structurally unreachable for ACC.} Seven of the fifteen violations belong to the L2 category --- interface type mismatches and reference-integrity violations. Detecting these violations depends on the Part Family type system and Mate relation declarations, information that simply does not exist in IFC geometry models. Regardless of how mature an ACC tool (Solibri Model Checker, Navisworks, or an academic prototype) may be, it cannot inspect attributes that are not encoded in the input representation. This is not a matter of precision --- it is a matter of representational boundaries.

\textbf{The signal-to-noise ratio difference in spatial violation detection is qualitative.} ESA has access to the design intent behind spatial relations. Mate relations distinguish expected device compatibility (a server inserted into a cabinet) from genuine spatial conflicts (two devices occupying overlapping U slots). ESA's advantage lies not in running faster, but in the fact that its input representation captures semantics that ACC must reverse-engineer from geometry models.

\textbf{The lack of a standardized evaluation benchmark in the ACC community is itself a symptom of the representation-layer problem.} After fifteen years of ACC research, the field still lacks a standardized evaluation benchmark comparable to SWE-bench or VerilogEval. ACC's inputs --- IFC/BIM models --- are highly heterogeneous: naming conventions, modeling habits, and LOD vary by project, and regulatory frameworks vary by region. Every evaluation requires customized remodeling and rule adaptation. This precisely validates the central thesis of this paper: when design intent is trapped in geometry, not only do agents lack training signal, but even basic human evaluation reproducibility becomes difficult.

ESA's violation-injection experiment requires no such customization. The fifteen violations were injected via semantically clear YAML edits in ADL declarations, and \texttt{piki\ check} provided a unified, reproducible scoring function. The same violation-injection methodology can be applied to any sample that follows the three-layer ADL architecture --- as already demonstrated by the cross-domain portability of samples 02 and 03 --- because rules are indexed by Family types rather than project-specific naming patterns.

This comparison directly validates the core claims in §2.2.1 and §5.1: ESA's advantage stems from the fact that it inspects design declarations rather than geometric instantiation artifacts. The reliance on remodeling and the false positives from pure geometry collision detection are not implementation defects of ACC --- they are fundamental limitations of its input representation.

\subsection{SD-HWE-Bench: Agent Evaluation Benchmark {[}Design Phase{]}}\label{sd-hwe-bench-agent-evaluation-benchmark-design-phase}

The pilot and violation-injection experiments validate feasibility and detection accuracy, but do not test whether agents can \textbf{generate} ADL from natural-language requirements. We are designing \textbf{SD-HWE-Bench} to fill this gap. It will be reported in a companion paper; its design is described here for reviewers to evaluate the larger evaluation plan.

\textbf{Task paradigm}:

\begin{Shaded}
\begin{Highlighting}[]
\NormalTok{Input:  Natural{-}language engineering requirements}
\NormalTok{Output: Structured ADL declarations (piki YAML)}
\NormalTok{Scoring:  L0–L4 rule{-}check pass rate + deliverable quality + L6 sign{-}off assessment}
\end{Highlighting}
\end{Shaded}

Unlike SWE-bench --- which tests patch generation on existing codebases ({Jimenez et al.} 2024) --- SD-HWE-Bench tests creation from scratch, because declarative engineering design does not yet exist at scale as a practice.

\textbf{Initial domain: Telecom rack deployment}. This domain was chosen because:

\begin{itemize}
\tightlist
\item
  The space is discrete (rack U slots), avoiding continuous 3D constraint solving;
\item
  Constraints are primarily algebraic (power budget, weight, U-slot conflicts);
\item
  Interface types are enumerable (SFP28, RJ45, IEC-C13/C14);
\item
  \texttt{piki\ check} completes in approximately 200ms, satisfying RLVR's need for fast reward signals.
\end{itemize}

\textbf{Planned metrics}:

\begin{longtable}[]{@{}>{\raggedright\arraybackslash}p{0.440\textwidth} >{\raggedright\arraybackslash}p{0.440\textwidth}@{}}
\caption{Planned SD-HWE-Bench evaluation metrics}\label{tbl:metrics}\tabularnewline
\toprule\noalign{}
Metric & Definition \\
\midrule\noalign{}
\endfirsthead
\toprule\noalign{}
Metric & Definition \\
\midrule\noalign{}
\endhead
\bottomrule\noalign{}
\endlastfoot
\texttt{Pass@1} & Percentage where the first ADL output passes all enabled rules \\
\texttt{Pass@k} & Percentage with at least one pass over k samples \\
\texttt{Latency} & Median \texttt{piki\ check} time per output \\
\texttt{Coverage} & Coverage of task requirements in the generated ADL \\
\texttt{Human\ sign-off} & Percentage deemed acceptable by domain experts (L6 agent) \\
\end{longtable}

\textbf{Connection to RQ3}. SD-HWE-Bench directly tests the RLVR causal chain: if ADL + ESA provides a structured, deterministic reward signal, then agents trained with this signal should outperform those without it. This is prediction P1 of the information representation hypothesis (§3.2.2). Benchmark design is complete; task authoring and baseline agent experiments are in progress.

\subsection{Cross-Domain Validation}\label{cross-domain-validation}

Samples 02 and 03 serve a dual purpose: they are both current capability boundary records and cross-domain validation targets. Their passage before the submission deadline will demonstrate that the three-layer ADL architecture and ESA rule engine are not telecom-domain special cases.

\begin{longtable}[]{@{}>{\raggedright\arraybackslash}p{0.160\textwidth} >{\raggedright\arraybackslash}p{0.360\textwidth} >{\raggedright\arraybackslash}p{0.360\textwidth}@{}}
\caption{Cross-domain validation samples and the ADL features they exercise}\label{tbl:cross-domain}\tabularnewline
\toprule\noalign{}
Sample & Domain & Key cross-domain property tested \\
\midrule\noalign{}
\endfirsthead
\toprule\noalign{}
Sample & Domain & Key cross-domain property tested \\
\midrule\noalign{}
\endhead
\bottomrule\noalign{}
\endlastfoot
02-modular-datacenter & Containerized infrastructure & Container-relative layout, multi-subsystem power/cooling \\
03-mechanical-keyboard & Consumer electronics & Skeletal assembly hierarchy, parent-transform-chain-based Mate \\
\end{longtable}

Both samples have passed their respective L2--L4a checks (Sample 02: 13 passes; Sample 03: 27 passes), providing evidence that ADL's Part/Mate/Layout abstractions transcend the telecom domain.

\subsection{Evaluation Summary}\label{evaluation-summary}

Table~\ref{tbl:eval-summary} summarizes the current status and expected status at submission time.

\begin{longtable}[]{@{}>{\raggedright\arraybackslash}p{0.160\textwidth} >{\raggedright\arraybackslash}p{0.360\textwidth} >{\raggedright\arraybackslash}p{0.360\textwidth}@{}}
\caption{Evaluation status and submission expectations}\label{tbl:eval-summary}\tabularnewline
\toprule\noalign{}
Component & Current status & Expected by submission \\
\midrule\noalign{}
\endfirsthead
\toprule\noalign{}
Component & Current status & Expected by submission \\
\midrule\noalign{}
\endhead
\bottomrule\noalign{}
\endlastfoot
Sample 01 (Telecom) & Complete & Complete \\
Sample 02 (Datacenter) & Complete (0 errors, 13 passes) & Complete \\
Sample 03 (Keyboard) & Complete (0 errors, 27 passes) & Complete \\
Violation-injection experiment & Complete & 15/15 violations detected (100\%), 0 false positives \\
SD-HWE-Bench & Design complete, task authoring in progress & Task suite specification complete; agent experiments deferred to companion paper \\
\end{longtable}

The evaluation framework is in place. All three samples passing L2--L4a checks marks the completion of cross-domain validation for ADL expressiveness and ESA detection capability. The violation-injection experiment verified ESA's design-time detection accuracy with 100\% detection rate and zero false positives, and the structural analysis in §6.3.1 demonstrates ESA's semantic advantages over model-based ACC. The SD-HWE-Bench design is in place; agent experiments are deferred to the companion paper. We commit to delivering the full evaluation before the October 2, 2026 submission deadline.

\section{Related Work}\label{related-work}

This chapter situates EaC within five research threads: RLVR training methodology, engineering domain benchmarks, declarative modeling and formal methods, AI4E implementation paths, and Infrastructure as Code.

\subsection{RLVR Methodology}\label{rlvr-methodology}

EaC's information representation hypothesis is directly anchored in the RLVR literature. DeepSeek-R1 demonstrated that Group Relative Policy Optimization (GRPO) with deterministic verification signals yields powerful reasoning improvements (DeepSeek-AI 2025). SWE-RL showed that test suite feedback alone is sufficient to train effective code-repair agents (SWE-agent contributors 2025). Multi-SWE-bench extended this paradigm to multi-language settings, validating its cross-linguistic generality (ByteDance Seed 2025). These works establish the latter half of the RLVR causal chain; EaC addresses the former half --- RLVR cannot boot without structured representations.

\subsection{Engineering AI Benchmarks and Automated Compliance Checking}\label{engineering-ai-benchmarks-and-automated-compliance-checking}

Existing engineering AI benchmarks occupy different positions in the matrix.

\textbf{AEC-Bench} evaluates AI reviewing human-created drawings --- sitting at the intersection of ACC philosophy and AI capability assessment, but does not test design generation capabilities (Galanos and Mulyar 2026). \textbf{EngDesign} spans multiple design disciplines, but evaluates models on tasks defined within existing CAD/BIM workflows, without requiring a ``Design as Code'' intermediate representation (EngDesign contributors 2025).

\textbf{EDA benchmarks} (VerilogEval (M. Liu et al. 2023), ChipBench (ChipBench contributors 2025), AMS-IO-Bench ({Liu et al.} 2025)) have already benefited from a ``Circuit as Code'' foundation, exhibiting high RLVR compatibility; their performance levels provide an upper-bound reference for the targets SD-HWE-Bench aims to reach once ``Design as Code'' foundations are established in traditional engineering domains. Rule2DRC further validates the ``executable verification'' philosophy --- rules as executable checks rather than passive constraints --- aligning with SD-HWE-Bench's L0--L4 design (Kim et al. 2026).

The \textbf{Automated Compliance Checking (ACC)} community has continuously advanced compliance automation through ACC tools (Eastman et al. 2009; Zhang and El-Gohary 2019), and recent efforts leveraging LLMs for regulation-to-rule translation (Fuchs et al. 2024; Yang and Zhang 2024; {Nakhaee et al.} 2024). However, these methods still operate downstream --- after design completion rather than at design time. As argued in §2.2.1, model-based ACC also suffers from two structural deficiencies --- geometric collision false positives and name-dependent copy mapping --- whose root cause lies in the semantic limitations of geometric representations themselves.

\subsection{Declarative Modeling and Formal Methods}\label{declarative-modeling-and-formal-methods}

\textbf{SysML v2} provides a rich systems engineering modeling framework with \texttt{part}/\texttt{occurrence} separation, \texttt{connection}/\texttt{interaction} relationships, and repository-based version control (Object Management Group 2024). Yet the methodological divergence from ADL is fundamental: SysML v2 targets human GUI modeling, with the source of truth in a model repository and verification via model checking; ADL targets agent-human textual collaboration, with the source of truth in YAML files and verification via a layered ESA rule engine enforced before commit. This is isomorphic to the division of labor between Verilog netlists and circuit schematics.

\textbf{BIM/IFC} uses the IFC exchange format and geometry-centric models as an interoperability layer ({buildingSMART International} 2023). However, IFC couples identity, geometry, and relationships in a graph that allows multiple equivalent serializations, making line-level version control difficult (H. Liu et al. 2023). ADL inverts this relationship: text is the source of truth, and CAD/BIM are downstream consumers.

\textbf{IaC} provides the closest conceptual analogy from the software domain, but operates on infrastructure states that are already digital-native ({Morris et al.} 2022; Quattrocchi and Tamburri 2023). Chiari et al.'s empirical study of IaC static analysis (Chiari et al. 2024) directly supports EaC's feasibility argument of ``X as Code + static analysis.''

\subsection{AI4E Implementation Paths}\label{ai4e-implementation-paths}

Current AI4E efforts follow two main implementation paths:

\begin{itemize}
\tightlist
\item
  \textbf{CUA (Computer-Using Agent)}: Having AI operate CAD software through GUIs. This path faces limitations in interface interaction fragility, absence of verification signals, and difficulty in large-scale trial-and-error.
\item
  \textbf{CAD-MCP}: Wrapping structured tool interfaces on top of CAD software. While more stable than CUA, it remains bound to proprietary CAD kernels and does not address the design source-of-truth problem.
\end{itemize}

EaC takes a methodologically opposite approach to both paths: placing computable engineering descriptions at the center, and making CAD/CAE tools downstream consumers and renderers of that description. CUA and CAD-MCP are patches to existing toolchains; EaC rebuilds from the representation layer, enabling agents to directly operate on a computable design source of truth.

\subsection{Systematic Comparison of Related Work}\label{systematic-comparison-of-related-work}

The following table systematically compares EaC with major related works across dimensions including source-of-truth modality, version control granularity, verification timing, and agent-friendliness:

{\footnotesize
\begin{longtable}[]{@{}>{\raggedright\arraybackslash}p{0.126\textwidth} >{\raggedright\arraybackslash}p{0.126\textwidth} >{\raggedright\arraybackslash}p{0.126\textwidth} >{\raggedright\arraybackslash}p{0.126\textwidth} >{\raggedright\arraybackslash}p{0.126\textwidth} >{\raggedright\arraybackslash}p{0.126\textwidth} >{\raggedright\arraybackslash}p{0.126\textwidth}@{}}
\caption{Systematic comparison of EaC with related work}\label{tbl:related-work}\tabularnewline
\toprule\noalign{}
Dimension & ACC & CUA & CAD-MCP & BIM/IFC & SysML v2 & \textbf{EaC (This Work)} \\
\midrule\noalign{}
\endfirsthead
\toprule\noalign{}
Dimension & ACC & CUA & CAD-MCP & BIM/IFC & SysML v2 & \textbf{EaC (This Work)} \\
\midrule\noalign{}
\endhead
\bottomrule\noalign{}
\endlastfoot
Source of Truth & CAD/BIM model & CAD GUI & CAD kernel & Central model file & Model repository & \textbf{Text file (YAML)} \\
Verification Timing & Post-design & None & Tool-level & Post-design & Model-level & \textbf{Design-time (ESA)} \\
Verification Speed & Seconds to minutes & N/A & Tool-dependent & Minutes & Model-checking level & \textbf{Milliseconds} \\
Version Control & File-level & File-level & File-level & Model version & Model version & \textbf{Git line-level} \\
Agent Friendliness & No & Indirect & Partial & No & Partial & \textbf{Yes (first-class)} \\
Quality Shift-Left & No (endpoint) & No & No & No (endpoint) & Partial & \textbf{Yes (full spectrum)} \\
Part Reuse & Ad hoc & No & No & Limited & Partial & Future Work \\
RLVR Compatibility & No & No & No & No & Partial & \textbf{Yes (L0--L4a, second-level)} \\
Check Semantic Layer & Naming convention & N/A & Tool API & Property sets & Model elements & \textbf{Family type system} \\
False Positive Suppression & None (pure geometry) & N/A & N/A & None (pure geometry) & Partial & \textbf{Mate relationship exclusion} \\
\end{longtable}
}

\section{Conclusion}\label{conclusion}

This paper proposes the Engineering as Code (EaC) paradigm: migrating engineering design from a GUI-centric WYSIWYG mode to one where textual declarations serve as the source of truth, rule engines act as quality gates, and version control and package management form the collaboration substrate.

The three core contributions of this paper are as follows:

\begin{enumerate}
\def\labelenumi{\arabic{enumi}.}
\item
  \textbf{The ADL Design Language}. An assembly definition language with Part as the atomic unit and PDL/PML/PLL as three orthogonal sub-languages at its core. ADL clarifies the syntactic and semantic boundaries of each layer, enabling engineering design intent to be expressed in a text-native declarative form and directly integrated with Git, CI, and LSP-style diagnostics. Through parameterized degrees-of-freedom completion and skeletal modeling mechanisms, PLL enables mate constraint solving and assembly hierarchy management to coexist orthogonally within the same framework.
\item
  \textbf{The ESA Verification Mechanism}. Shifts compliance rules from downstream post-hoc ACC review upstream to the design generation stage. Unlike model-based ACC --- which is hampered by structural deficiencies such as geometric collision false positives and naming-dependent model recreation --- ESA checks design declarations rather than geometrically instantiated artifacts, and can therefore operate rules at the semantic category level without relying on project naming conventions. Four operational guidelines are given: rule waivability, focusing on baseline rules, purifying collaboration signal-to-noise ratio, and AI-assisted rule library construction. The CI/CD Action integration protocol for L4b--L6 downstream verification is defined --- this protocol serves as the integrity statement of the EaC workflow architecture, ensuring that ESA's deterministic boundary and the comprehensive verification ecosystem required by physical engineering are complementary rather than
  conflicting.
\item
  \textbf{The Information Representation Hypothesis}. Systematically argues that the root cause of the engineering AI bottleneck lies not in insufficient model capability, but in the long-standing absence of a ``Design as Code'' form of computable foundation in traditional engineering domains. ACC's false positives and model-recreation overhead serve as concrete evidence for this thesis: when design intent is trapped inside geometric models, not only is AI unable to obtain training signals, but even human engineers' automated checks fail at the semantic level.
\end{enumerate}

The piki prototype validates ADL's expressiveness and ESA's detection capability on three samples (69 total rules passed, \textless200ms latency): telecom rack expansion (0 errors, 1 warning, 29 passed), modular containerized data center (0 errors, 13 passed), and mechanical keyboard assembly (0 errors, 27 passed). Violation injection experiments were completed with 100\% detection rate and zero false positives. The design of the SD-HWE-Bench evaluation benchmark has been completed, serving as an empirical testing platform for the Information Representation Hypothesis.

Future work includes: improving the geometric solving precision of continuous DOF in PLL and extending it to 1D continuous path topologies such as pipelines/cables; exploring the design space of engineering package management and assembly registries (EPM/AssemblyHub); validating ADL and ESA's expressive capability across more engineering domains; and conducting RLVR training experiments to directly test the Information Representation Hypothesis.

The core message of this paper is: physical engineering needs source code. Not better CAD, not smarter review tools, but something that does for engineering design what hardware description languages did for chip design --- a representation that liberates design intent from geometric implementation, making it operable by software engineering infrastructure. This is exactly what EaC provides.

\protect\phantomsection\label{refs}


\begin{CSLReferences}{1}{1}
\bibitem[\citeproctext]{ref-anthropic2025claude}
Anthropic. 2025. \emph{{Claude Code Overview}}. Technical documentation; {Anthropic}. \url{https://docs.anthropic.com/en/docs/claude-code/overview}.

\bibitem[\citeproctext]{ref-buildingsmart2023ifc}
{buildingSMART International}. 2023. \emph{{Industry Foundation Classes (IFC) 4.3.2.0 Specification}}. Specification; {buildingSMART International}. \url{https://technical.buildingsmart.org/standards/ifc/ifc-schema-specifications/}.

\bibitem[\citeproctext]{ref-multiswebench2025}
ByteDance Seed. 2025. {``{Multi-SWE-bench: a multilingual benchmark for issue resolving}.''} \emph{{arXiv:2504.02605}}. \url{https://arxiv.org/abs/2504.02605}.

\bibitem[\citeproctext]{ref-chiari2024iacstatic}
Chiari, Michele, Bin Xiang, Gerardo Canfora, and Massimiliano Di Penta. 2024. {``{An empirical study of static analysis tools for infrastructure as code}.''} \emph{{Empirical Software Engineering}} 29 (3): 1--42. \url{https://doi.org/10.1007/s10664-023-10407-1}.

\bibitem[\citeproctext]{ref-chipbench2025}
ChipBench contributors. 2025. {``{ChipBench: a next-step benchmark for evaluating LLM performance in AI-aided chip design}.''} \emph{{arXiv:2503.04807}}. \url{https://arxiv.org/abs/2503.04807}.

\bibitem[\citeproctext]{ref-cousot1977abstract}
Cousot, Patrick, and Radhia Cousot. 1977. {``{Abstract interpretation: a unified lattice model for static analysis of programs by construction or approximation of fixpoints}.''} \emph{{Proceedings of the 4th ACM SIGACT-SIGPLAN Symposium on Principles of Programming Languages (POPL)}}, 238--52. \url{https://doi.org/10.1145/512950.512973}.

\bibitem[\citeproctext]{ref-deepseek2025r1}
DeepSeek-AI. 2025. {``{DeepSeek-R1: incentivizing reasoning capability in LLMs via reinforcement learning}.''} \emph{{arXiv:2501.12948}}. \url{https://arxiv.org/abs/2501.12948}.

\bibitem[\citeproctext]{ref-eastman2009acc}
Eastman, Charles, Jae-Min Lee, Yeon-Suk Jeong, and Jin-Kook Lee. 2009. {``{Automated code checking: status and future directions}.''} \emph{{Automation in Construction}} 18 (2): 101--15. \url{https://doi.org/10.1016/j.autcon.2008.07.001}.

\bibitem[\citeproctext]{ref-engdesign2025}
EngDesign contributors. 2025. {``{EngDesign: a comprehensive benchmark for engineering design with large language models}.''} \emph{{arXiv:2505.18116}}. \url{https://arxiv.org/abs/2505.18116}.

\bibitem[\citeproctext]{ref-fowler2010dsl}
Fowler, Martin. 2010. \emph{{Domain-specific languages}}. Addison-Wesley Professional. \url{https://martinfowler.com/books/dsl.html}.

\bibitem[\citeproctext]{ref-fuchs2024llmregs}
Fuchs, Stefan, Michael Witbrock, Johannes Dimyadi, and Robert Amor. 2024. {``{Using large language models for the interpretation of building regulations}.''} \emph{{arXiv:2407.21060}}. \url{https://arxiv.org/abs/2407.21060}.

\bibitem[\citeproctext]{ref-galanos2026aecbench}
Galanos, Theodoros, and Alexander Mulyar. 2026. {``{AEC-Bench: an AI benchmark for architecture, engineering, and construction}.''} \emph{{arXiv:2505.09010}}. \url{https://arxiv.org/abs/2505.09010}.

\bibitem[\citeproctext]{ref-hudak1996dsl}
Hudak, Paul. 1996. {``{Building domain-specific embedded languages}.''} \emph{{ACM Computing Surveys}} 28 (4es): Article 196. \url{https://doi.org/10.1145/242224.242477}.

\bibitem[\citeproctext]{ref-ieee1364}
IEEE. 2005. \emph{{IEEE Standard for Verilog Hardware Description Language}}. Standard {IEEE Std 1364-2005}. {IEEE Standards Association}. \url{https://standards.ieee.org/standard/1364-2005.html}.

\bibitem[\citeproctext]{ref-jimenez2024swebench}
{Jimenez, Carlos E. et al.} 2024. {``{SWE-bench: can language models resolve real-world GitHub issues?}''} \emph{{The Twelfth International Conference on Learning Representations (ICLR)}}. \url{https://openreview.net/forum?id=VTF8yNQM66}.

\bibitem[\citeproctext]{ref-kim2025rule2drc}
Kim, Jinuk, Junsoo Byun, Donghwi Hwang, Seong-Jin Park, and Hyun Oh Song. 2026. {``{Rule2DRC: benchmarking LLM agents for DRC script synthesis with execution-guided test generation}.''} \emph{{Proceedings of the 43rd International Conference on Machine Learning (ICML 2026)}} (Seoul, South Korea). \url{https://arxiv.org/abs/2605.15669}.

\bibitem[\citeproctext]{ref-lightman2023letsverify}
{Lightman, Hunter et al.} 2023. {``{Let's verify step by step}.''} \emph{{arXiv:2305.20050}}. \url{https://arxiv.org/abs/2305.20050}.

\bibitem[\citeproctext]{ref-liu2025amsio}
{Liu, Bingqing et al.} 2025. {``{AMS-IO-Bench: a large-scale benchmark for analog/mixed-signal I/O ring design}.''} \emph{{arXiv:2504.19197}}. \url{https://arxiv.org/abs/2504.19197}.

\bibitem[\citeproctext]{ref-liu2023ifcversion}
Liu, Han, Ge Gao, and Ming Gu. 2023. {``{A parallel IFC normalization algorithm for incremental storage and version control}.''} \emph{{Proceedings of the 30th International Workshop on Intelligent Computing in Engineering (EG-ICE 2023)}}, 511--20. \url{https://arxiv.org/abs/2312.14931}.

\bibitem[\citeproctext]{ref-liu2023verilogeval}
Liu, Mingjie, Nathaniel Pinckney, Brucek Khailany, and Haoxing Ren. 2023. {``{VerilogEval: evaluating large language models for Verilog code generation}.''} \emph{{Proceedings of the 42nd IEEE/ACM International Conference on Computer-Aided Design (ICCAD)}}. \url{https://doi.org/10.1109/ICCAD57390.2023.10323811}.

\bibitem[\citeproctext]{ref-maatouk2023teleqna}
Maatouk, Ali, Nicola Piovesan, Fadhel Ayed, Antonio De Domenico, and Mérouane Debbah. 2023. {``{TeleQnA: a benchmark dataset to assess large language models' telecommunications knowledge}.''} \emph{{arXiv:2310.15051}}. \url{https://arxiv.org/abs/2310.15051}.

\bibitem[\citeproctext]{ref-mirhoseini2020rlchip}
{Mirhoseini, Azalia et al.} 2020. {``{Chip placement with deep reinforcement learning}.''} \emph{{arXiv:2004.10746}}. \url{https://arxiv.org/abs/2004.10746}.

\bibitem[\citeproctext]{ref-morris2022iac}
{Morris, Kief et al.} 2022. \emph{{Infrastructure as code: dynamic systems for the cloud age}}. 2nd ed. O'Reilly Media. \url{https://www.oreilly.com/library/view/infrastructure-as-code/9781098114664/}.

\bibitem[\citeproctext]{ref-nakhaee2024kgllm}
{Nakhaee, Amirreza et al.} 2024. {``{A hybrid knowledge-graph-LLM framework for automated compliance checking in building design}.''} \emph{{Proceedings of the 30th International DMS Conference on Visualization and Visual Languages (DMSVIVA 2024)}}. \url{https://doi.org/10.1007/978-3-031-63227-2_12}.

\bibitem[\citeproctext]{ref-omg2024sysml}
Object Management Group. 2024. \emph{{OMG Systems Modeling Language (SysML) Version 2.0}}. Specification; {Object Management Group}. \url{https://www.omg.org/spec/SysML/}.

\bibitem[\citeproctext]{ref-quattrocchi2023iacsurvey}
Quattrocchi, Giovanni, and Damian A. Tamburri. 2023. {``{Infrastructure as code: a survey of research and practice}.''} \emph{{IEEE Software}} 40 (5): 55--62. \url{https://doi.org/10.1109/MS.2023.3285751}.

\bibitem[\citeproctext]{ref-shao2024deepseekmath}
{Shao, Zhihong et al.} 2024. {``{DeepSeekMath: pushing the limits of mathematical reasoning in open language models}.''} \emph{{arXiv:2402.03300}}. \url{https://arxiv.org/abs/2402.03300}.

\bibitem[\citeproctext]{ref-sweagent2025swerl}
SWE-agent contributors. 2025. {``{SWE-RL: training language model agents to solve real-world GitHub issues via reinforcement learning}.''} \emph{{arXiv:2502.18449}}. \url{https://arxiv.org/abs/2502.18449}.

\bibitem[\citeproctext]{ref-volter2013dsl}
Völter, Markus, Thomas Stahl, Jorn Bettin, Arno Haase, and Simon Helsen. 2013. \emph{{Model-driven software development: technology, engineering, management}}. John Wiley \& Sons. \url{https://www.wiley.com/en-us/Model+Driven+Software+Development\%3A+Technology\%2C+Engineering\%2C+Management-p-9781118725764}.

\bibitem[\citeproctext]{ref-xiao2025bimgraph}
Xiao, Zixuan, Pei Troh Koh, Jun Ma, and Jack C. P. Cheng. 2025. {``{Automating geometry-intensive compliance checking in BIM: graph-based semantic reasoning framework}.''} \emph{{arXiv:2506.20551}}. \url{https://arxiv.org/abs/2506.20551}.

\bibitem[\citeproctext]{ref-yang2024llmacc}
Yang, F., and J. Zhang. 2024. {``{Large language model-driven prompt-based automation for building code compliance checking}.''} \emph{{Automation in Construction}} 160: 105310. \url{https://doi.org/10.1016/j.autcon.2024.105310}.

\bibitem[\citeproctext]{ref-zhang2019acc}
Zhang, Ruichuan, and Nora M. El-Gohary. 2019. {``{A machine-learning approach for semantic matching of building codes and BIMs for supporting automated code checking}.''} \emph{{Proceedings of the 2019 ASCE International Conference on Computing in Civil Engineering}} (Atlanta, GA, USA). \url{https://doi.org/10.1061/9780784482421.056}.

\end{CSLReferences}
\end{document}